\documentclass[preprint,12pt,authoryear]{elsarticle}
\usepackage{hyperref}

\journal{Journal of \LaTeX\ Templates}

\usepackage{newtxtext,newtxmath}
\usepackage[T1]{fontenc}
\usepackage{ae,aecompl}
\usepackage{cleveref}

\usepackage{graphicx}	
\usepackage{amsmath}	
\usepackage{amssymb}	

\usepackage{booktabs}
\usepackage{wasysym}
\usepackage{bm}
\usepackage{mathtools}
\usepackage{etoolbox}

\def\eg{{\it e.g.}}
\def\beq{\begin{equation}}
\def\eeq{\end{equation}}
\def\pl#1#2{{\frac{\partial #1}{\partial #2}}}
\def\dl#1#2{{\frac{d #1}{d #2}}}
\def\ibid{{\it ibid.\thinspace}}
\def\ie{{\it i.e.}}

\begin{document}

\begin{frontmatter}

\title{Fine-grained rim formation - high speed, kinetic dust aggregation in the early Solar System}

\author{Kurt Liffman}
\address{Centre for Astrophysics and Supercomputing, Swinburne University of Technology, Hawthorn, Victoria 3122, Australia}

\ead{kliffman@swin.edu.au}

\begin{abstract}
		Type 3 chondritic meteorites often contain chondrules and refractory inclusions that are coated with accretionary, fine-grained rims (FGRs). FGRs are of low porosity, were subject to centrally directed pressure, may contain high temperature products like microchondrules and there is a linear relationship between the rim thickness and the radius of the enclosed object.
		
		FGRs are thought to have formed by the gentle adhesion of dust onto the central object with the subsequent compression of this fluffy rim within the parent body. However, this model does not explain the low porosity, micro-chondrules and centralized pressure. This model also has difficulties explaining the linear relationship between rim thickness and object size including the existence of a non-zero constant in that linear relationship.
		
		We propose that FGRs formed by the relatively high-speed interaction between dust and the object, where high initial impact speed produced abrasion and, possibly, microchondrules. FGR formation occurred over a range of lower speeds aided by vacuum adhesion of fragments from the impacting dust particles. This model naturally produces the rim thickness linear relationship with non-zero constant, low porosity and centrally directed pressure. We call this process kinetic dust aggregation (KDA), which is another name for the aerosol deposition processes used in industry. KDA may be a tentative, part explanation of how dust aggregation occurs in protostellar disks on the pathway from dust to planets.
\end{abstract}

\begin{keyword}
CAIs, chondrules, fine-grained rims, accretionary rims, solar nebula, protostellar disks
\end{keyword}

\end{frontmatter}

	
	
	\section{Introduction}
	
\begin{figure}
        \includegraphics[width=\columnwidth]{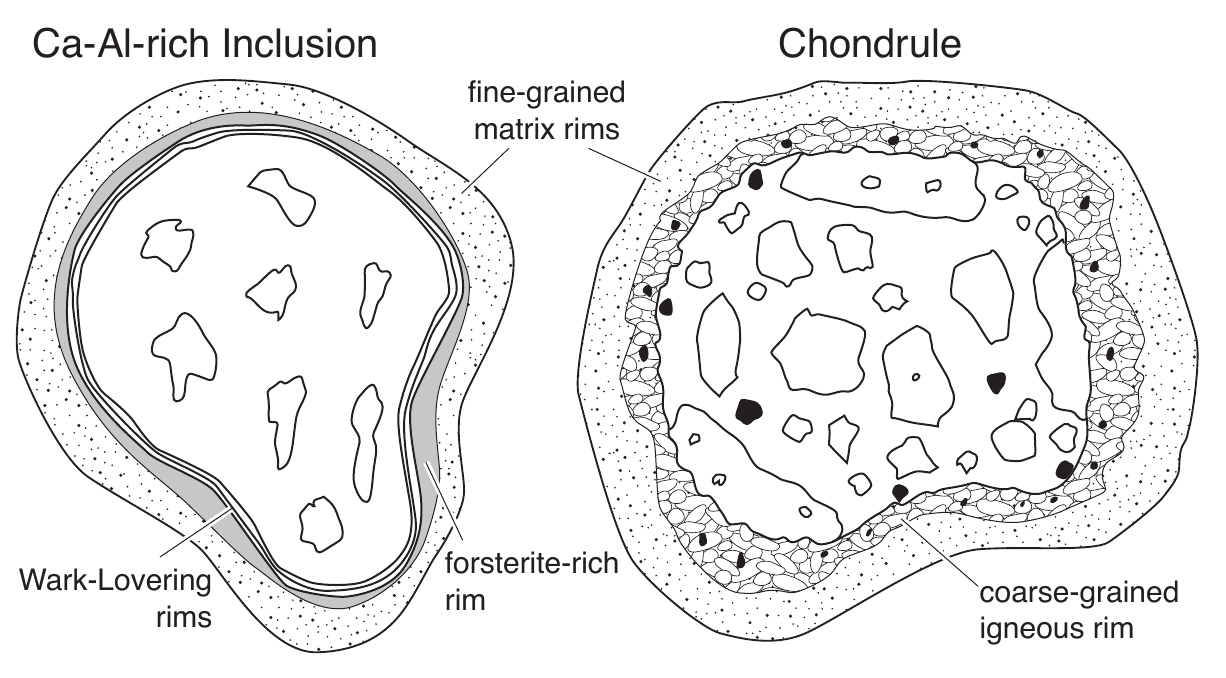}
		\caption{Fine Grained Rims (FGRs) are often found around chondrules, calcium aluminum inclusions (CAIs) and other objects within well preserved, primitive meteorites. FGRs always coat other rims such as Wark Lovering Rims found on CAIs and igneous rims on chondrules. Although, it should be noted that many chondrules don't have igneous rims, but they still have FGRs. FGRs often also contain microchondrules and fragments from the underlying chondrule. Figure obtained with permission from \cite{2005ASPC..341...15S}.}
		\label{fig:FGR}
	\end{figure}

	Fine grained rims (FGRs) are often found around chondrules and calcium aluminum inclusions in primitive chondrites of petrologic type 3 (\cite{1996cpd..conf..153M} and Figure \ref{fig:FGR}). The process that produced these rims is still debated, but most authors agree that FGRs were formed when dust adhered to the free floating chondrules, CAIs and other such objects prior to that object being incorporated into the parent body of the meteorite (\citet{1992GeCoA..56.2873M,2005ASPC..341..732C}). However, other authors have proposed that FGRs were formed due to parent body processes (\citet{2006GeCoA..70.1271T}), when the central object was, for example, coated with slurry-like material within the parent body (\citet{2014GeCoA.137...18T}).
	
	In this paper, we will use but modify the conventional idea that FGRs formed due to the adhesion of dust onto free floating objects. As we will discuss, the numerous physical properties of FGRs can be potentially understood within a theoretical framework of relatively high speed collisions between the central object and micron-sized dust particles under vacuum conditions.

	When all the evidence is considered regarding FGR formation, it is difficult to 
	account for FGR characteristics as a purely parent-body process. For example, 
	\cite{2016E&PSL.434..117L} and \cite{2018GeCoA.221..379H} examined presolar dust 
	grains in various carbonaceous chondrites and found that, under certain circumstances,  there were significant differences in the concentrations of presolar grains in FGRs relative to the chondrite 
	matrix thereby suggesting that FGRs were formed on the chondrules prior to 
	inclusion into the chondrite parent body. 
	
	In particular, a parent body/slurry process may have considerable difficulty in  
	accounting for the most unusual observed property of fine-grained rims: where the
	thickness of the rim, $\Delta r$, is proportional to the radius of the rimmed 
	object, $a_0$:

	\beq
	\Delta r \approx K a_0 + c \ ,
	\label{eq:Del_a=Ka_0}
	\eeq
	where $K$ and $c$ are constants (\cite{1989LPI....20..689M}, 
	\cite{1992GeCoA..56.2873M}, \cite{1993Metic..28..669S}, 
	\cite{1997LPI....28.1071P}, \cite{2005M&PS...40.1413G}, 
	\cite{2018E&PSL.481..201H}, \cite{2018M&PS...53.2470F}).
	
	\cite{1992GeCoA..56.2873M} show results for four different meteorites 
	 (Kivesvaara, Y-791198, Y-74662 and Murray), where the rim thickness was plotted as a function of chondrule diameter, the corrected slope $K$ for chondrule radius is in the range 
	 of 0.34 to 0.42, while $c$ ranges from 8 to 26 microns. Here we note that these 
	 observational results were obtained from two dimensional thin sections of the 
	 meteorites and were not corrected for the 2D/3D thin section size bias (\eg, 
	 \cite{2017M&PS...52..532C}). This bias tends to overestimate the rim thickness 
	 relative to the diameter of the central object.
	 
\begin{figure}
		\includegraphics[width=\columnwidth]{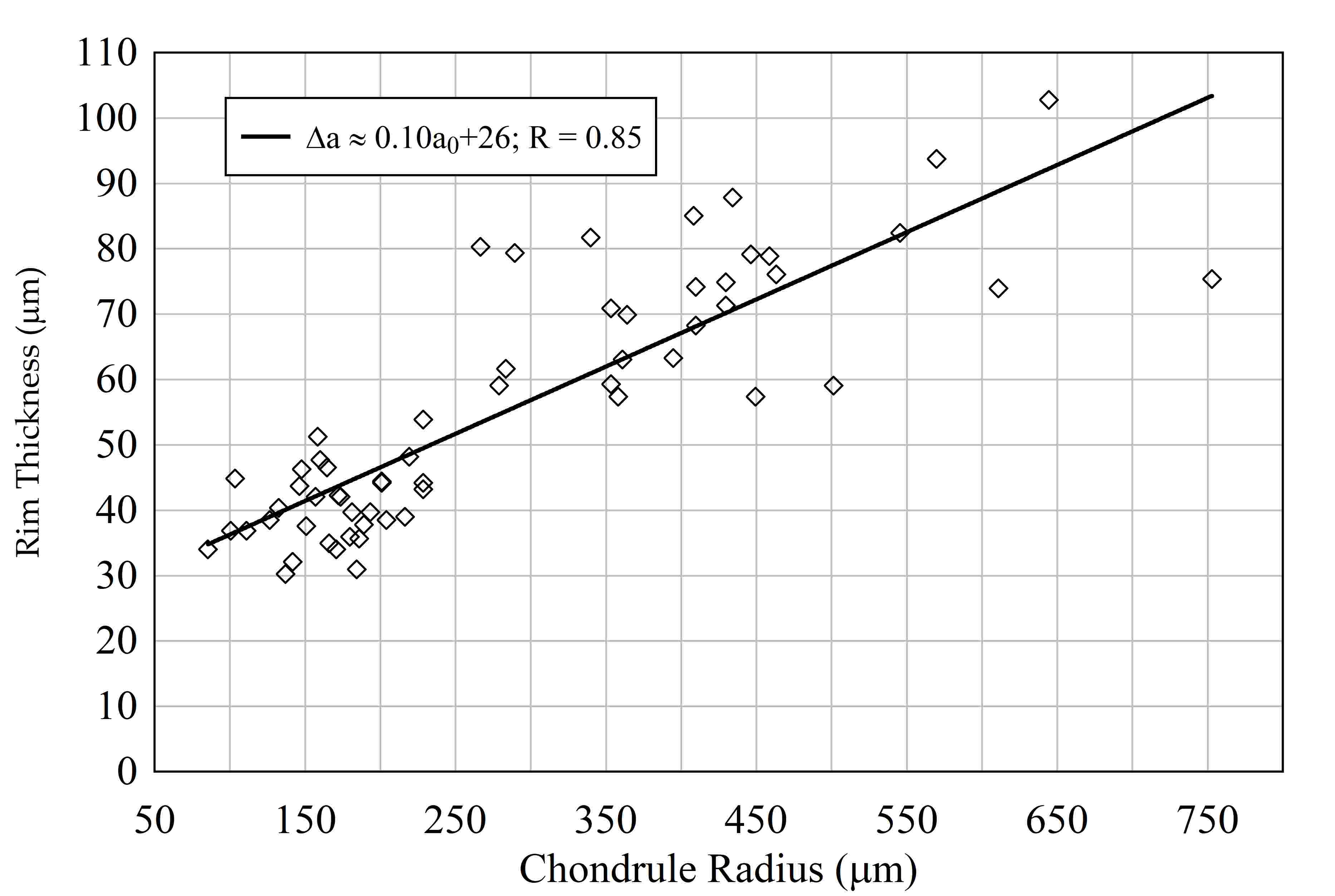}
		\caption{The thickness of the rim around chondrules is linearly related to the radius of the enclosed chondrule. Note that the relationship also includes a non zero constant. R is the correlation coefficient. Data obtained from \cite{2018E&PSL.481..201H} for the CM Murchison carbonaceous chondrite.}
		\label{fig:RIM}
	\end{figure}

	 Improvements in imaging technology have given researchers the capability
	(\cite{2018E&PSL.481..201H}) to use imaging techniques such as X-ray computed
	tomography (XCT) to examine the 3D morphology of Type I (\ie, FeO-poor) chondrules
	in the CM carbonaceous chondrite Murchison. These authors found that
	
	\beq
 	\Delta r \approx 0.1 a_0 + 26 \ ,
	\label{eq:Del_r}
	\eeq
	where the units are in microns (Figure \ref{fig:RIM}).
	
	 To date, all theoretical modelling of this phenomenon (\eg, \cite{1998Icar..134..180M}, \cite{2000Icar..143..106L} and \cite{2004Icar..168..484C}) has derived the equation $\Delta r \approx K a_0$ and has ignored the non-zero $c$ value. This was possibly because a non zero $c$ was considered an experimental artifact that would disappear with new and improved data. However, the latest observations suggest that non-zero $c$ values are real and is indicative of a process during FGR formation which affected the chondrules to an approximately constant depth, possibly at the chondrule surface, irrespective of the chondrule size.
	
	It may be relevant to note that surface erosion is a recurring theme in the FGR formation literature. Abrasion, erosion, reheating and possible melting of the chondrule surface appear to be markers of FGR formation (\eg, \cite{1986PolRe..41..222K}, \cite{1991Icar...91...76B}, \cite{1996cpd..conf..181K}, \cite{2012M&PS...47..142P} and \cite{2018GeCoA.221..379H}). The non-zero $c$ value might be associated with this erosive process. The presence of microchondrules and chondrule fragments mixed into the FGR with possibly related reheating of the chondrule surface is further evidence that a melt or fragmentation layer was involved with an energetic FGR formation process (\cite{1997GeCoA..61..463K}, \cite{2016M&PS...51..235B} and \cite{2016M&PS...51..884D}).
	
	Alternatively the non-zero $c$ value may simply represent the average compressed dust grain size of the layer of dust stuck to the chondrule surface. This interface layer of dust between the chondrule and the FGR should have the same thickness irrespective of the initial size of the chondrule - a result that is suggested by the simulation data given in Figure 7(a) of \cite{2019Icar..321...99X}.
	
	The most controversial aspect of FGRs is their low porosity and fine-grained nature. As discussed by \cite{2006GeCoA..70.1271T}, the standard formation theory of FGRs has great difficulty in explaining the compressed nature and low porosity of FGRs and is one of the reasons why parent body processes are still considered for FGR formation.
	
	 In standard FGR formation theory, FGRs were formed when micron-sized dust adhered to the central object, where the dust collision speeds were of order tens of centimetres per second or less (\cite{1993ApJ...407..806C}, \cite{1997ApJ...480..647D},  \cite{2004Icar..168..484C}, \cite{2008ARA&A..46...21B}). Although there is a subset of authors that consider higher dust collision speeds (e.g., \cite{1981Icar...48..460K}, \cite{1991Icar...91...76B}, \cite{1998Sci...280...62C} and \cite{2000Icar..143..106L}), most authors ignore collision speeds in excess of approximately a metre per second, because it is deduced from theoretical and experimental results that such speeds would not allow smooth, micron-sized dust particles to stick to each other and/or the chondrule/CAI surface (\cite{2010A&A...513A..56G}). The genesis of a FGR, according to standard theory, was an enclosed object surrounded by a porous, "fractal" layer of dust (\cite{2000Icar..143..138B}). The dust covered particle was then incorporated into a planetesimal, where compressional forces within the planetesimal converted the porous dust layer into a FGR.
	
	 Unfortunately, laboratory experiments have not been able to replicate the formation of FGRs via impact compression (\cite{2013Icar..225..558B}). Observations of stress in chondrule FGRs and surrounding matrix (\cite{2011NatGe...4..244B}) suggest that FGRs were exposed to a spherically symmetric stress field prior to the rimmed chondrule being incorporated into the parent body. The FGR formed while the chondrule was in free space and the initially porous dust particles were severely compacted during the rim formation process. \cite{2011NatGe...4..244B} concluded that the FGRs could not have formed within the parent body, because the FGR stress field and the stress field within the parent body appear to be completely independent.
	
	 So the most popular formation theory of FGRs is confronted by a contradiction: FGRs were formed while the chondrules/CAIs/ \&c. were in free space, but the assumed slow speed particle collisions produce a dust rim that is far too porous and cannot be converted into a FGR by internal pressures within the parent body.
	
	 One way to resolve this contradiction, is to note that some experiments do show that dust grains can stick to a surface if the collision speeds are in excess of a metre per second (\eg, \cite{2000ApJ...533..454P}, \cite{2007prpl.conf..783D} and \cite{2008ARA&A..46...21B}). This contradicts standard dust collision theory, however most analysis has ignored the possible role of fragmentation of the colliding dust particles. Unfortunately, ignoring dust fragmentation may have been a reasonable, but too simplistic approximation. For example, \cite{2008ARA&A..46...21B} were able to show experimentally that fragmentation can enhance adhesion for collision speeds in excess of 13 ms$^{-1}$. These authors suggest that fragmentation can absorb kinetic energy and, presumably, increase the dust particle surface area available for adhesion.
	
	  As discussed in \cite{ISI:000360783300001}, experimental experience in the fields of materials science and semiconductor manufacturing has shown that micron-sized ceramic particles can stick to a ceramic substrate and adhere to each other provided that the collision process occurs in a vacuum and the relative collision speeds are of order 100 metres per second. The resulting coatings have low porosity and are fine grained.

\begin{figure}
		\includegraphics[width=\columnwidth]{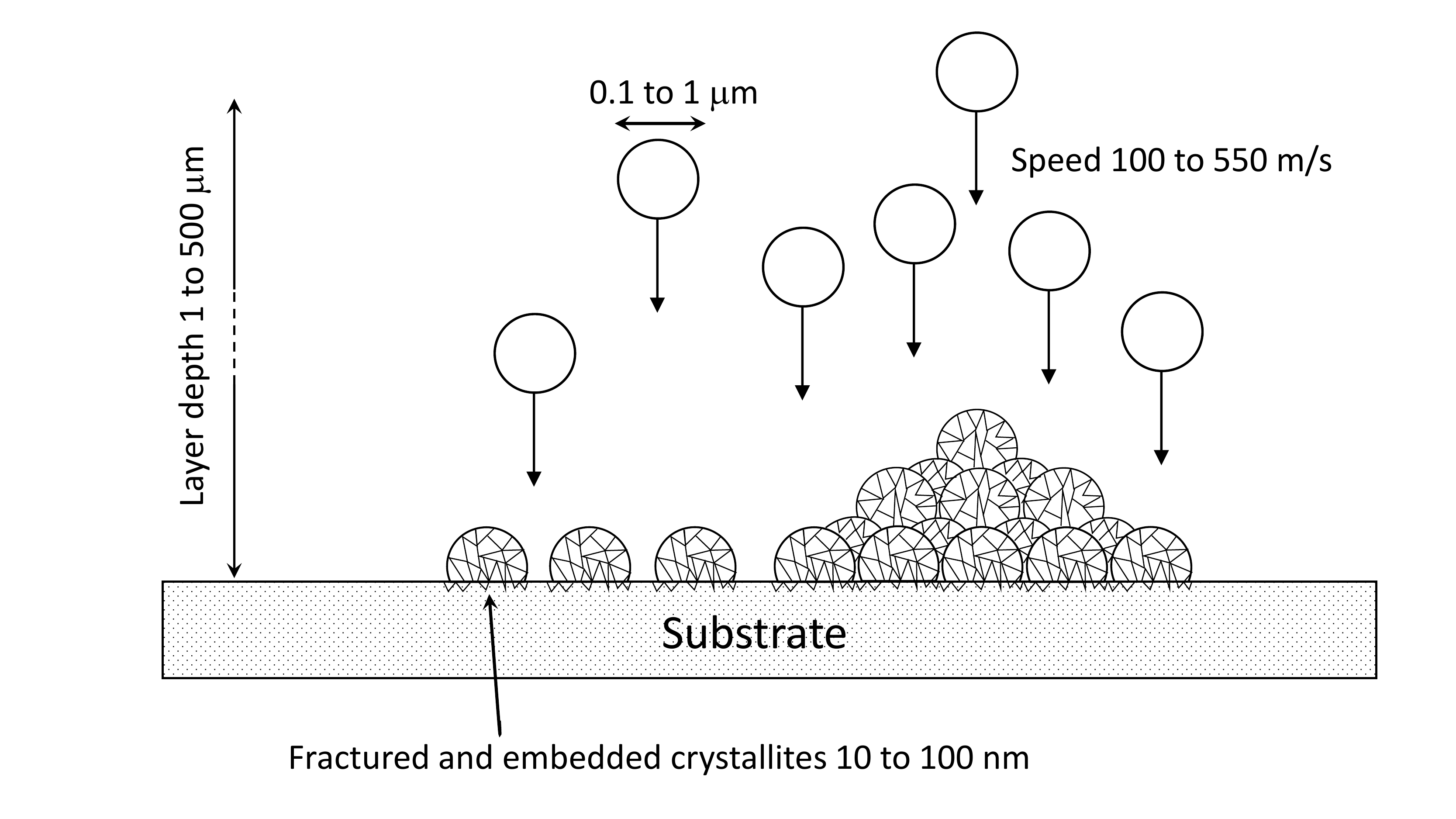}
		\caption{"Illustration of the fracturing and compaction of incident particles
			during aerosol deposition. The initial impact fractures and embeds the
			particles into the substrate. Subsequent impact causes further fracturing and
			compaction of the film." Caption and modified plot from \cite{ISI:000336118000040}.}
		\label{fig:Frac}
	\end{figure}

	This process is known by a number of different names, \eg, Vacuum Cold Spray (VCS), Vacuum Kinetic Spraying (VKS), Aerosol Deposition (AD) and Room Temperature Impact Consolidation (RTIC). In this paper, we will denote the process as "Kinetic Dust Aggregation" (KDA). A qualitative explanation for this phenomenon is shown in Figure \ref{fig:Frac} where the caption and figure have been copied from \cite{ISI:000336118000040}. To quote these authors:
	
	 "Figure \ref{fig:Frac} illustrates the currently accepted deposition mechanism. The film begins to form when incident particles impact, fracture, and embed into the substrate. This impact causes indentation and abrasion of the target area, giving rise to an increased surface area, which facilitates formation of a well-adhered anchor layer comprising 10 to 100 nm-sized particles. Subsequent impact compacts the underlying film and bonds the crystallites. Since only fracturing occurs, the fabricated film has the same crystal structure as the raw powder, which is composed of densely packed crystalline nano-particles held together by what is thought to be close-range mechanical and chemical interactions mediated by fracture and/or plastic deformation of the particles The ADM (Aerosol Deposition Method) can achieve a film density of 95\% of that of the bulk material"
	
    Note the mention of the nano-metre sized particles that are produced from fracturing the initial micron-sized particles. These nano-metre particles appear to be relatively sticky. A result that is consistent with \cite{2006JNR.....8..693R} who found that nano-metre sized particles still adhere even if the collision speeds approach 1 km s$^{-1}$.

\begin{figure}
		\includegraphics[width=\columnwidth]{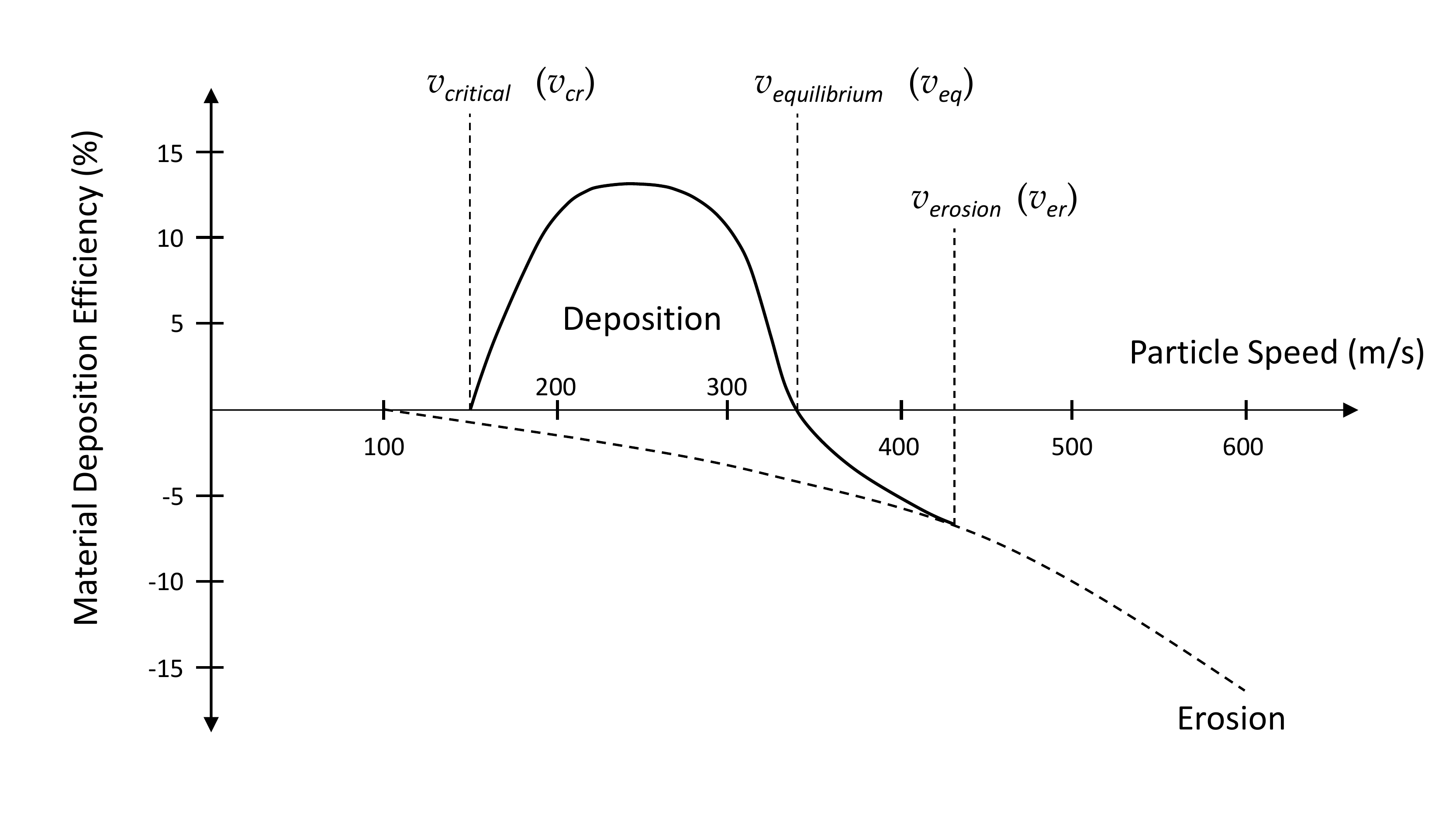}
		\caption{High speed collision between micron-sized particles can cause the particles to stick to each other or to a substrate. This Kinetic Dust Aggregation (DA) is due to the particles breaking up into smaller particles and the subsequent increase in particle surface area allowing greater adhesion and decreased porosity. There is a critical speed ($\varv_{critical}$ or $\varv_{cr}$) where if the speed of collision exceeds this critical speed then such adhesion can take place. At higher speeds, collisional erosion becomes competitive with adhesion and there is an equilibrium between adhesion and erosion rates. We call this speed $\varv_{equilibrium}$ ($\varv_{eq}$). At high speeds, the adhesive behaviour disappears until the interaction is fully erosive ($\varv_{erosion}$ or $\varv_{er}$). The dashed line shows the expected erosion if no adhesion were to take place. Modified plot from \cite{ISI:000360783300001}.}
		\label{fig:AD}
	\end{figure}

	From  \cite{ISI:000237812100010}, \cite{ISI:000255864400004} and \cite{ISI:000360783300001} we can draw Figure \ref{fig:AD}, which displays the observed, but still qualitative characteristics of the KDA process. As collision speeds increase, we reach a critical speed where the crystalline material starts to bond with the substrate. We denote this speed as the critical speed, $\varv_{cr}$ $\sim$ 150 ms$^{-1}$. In this case the critical speed is in excess of 100 ms$^{-1}$, but, as pointed out in \cite{2008ARA&A..46...21B}, meteoritic materials may have a much lower critical speed of near 10 ms$^{-1}$.  At speeds in excess of $\varv_{cr}$, the particles aggregate and form a bonded layer on the surface. This continues until we reach a collisional speed where the erosive affects of particle collisions balances the aggregation and the net increase in the layer size is zero. We denote this equilibrium speed by $\varv_{eq}$ ($\sim$ 300 ms$^{-1}$). Finally, as we continue to increase the collisional speed, the aggregation effect disappears and all the collisions produce an erosive effect. We denotes this speed as the erosion speed or $\varv_{er}$ ($\sim$ 500 ms$^{-1}$).
	
	Naturally, experiments have to be conducted with the appropriate materials and conditions to determine the particular critical and erosion speeds for meteoritical materials and whether FGRs and their associated microchondrules can be formed in the laboratory using these KDA techniques. It is a fond hope that this paper may prompt experiments along this research pathway.
	
	 In the next section, we consider the mathematical basis and fundamental assumptions underlying prior theories of FGR formation. We then use the KDA phenomenon to build upon these prior ideas to produce a new model for understanding the formation of FGRs.

	\section{Theory of FGR Formation}
\label{sec:FGR_theory}

In the first two subsections, we will consider the foundation state of FGR formation theory. In the third subsection we will outline the theory as developed in this study.

\subsection{Constant spatial mass density of dust}

Most of the basic mathematical theory for FGR formation can be found in \cite{1998Icar..134..180M} (see also the more general formulae in  \cite{2004Icar..168..484C}). Using this standard theory, in \ref{sec:const_density} we derive the formula for the thickness of an accretion rim for a macroscopic particle moving through a dusty gas of constant spatial dust mass density:
\beq
\Delta r = a_p(0)(e^{t/t_a} - 1) \ ,
\label{eq:Delta_r_th2}
\eeq
here $a_p(0)$ is the radius of the macroscopic particle at time $t=0$. The accretion time-scale, $t_a$, has the form

\beq
\begin{split}
t_a & = \frac{4{\rho}_pa_p(0)}{Q\rho_{sd} \varv_{pg}(0)} \\
& \approx 3,800 \ {\rm yr} \
\frac{({\rho}_p/3 \ {\rm g cm}^{-3}) \left(a_p(0)/1 \ {\rm mm} \right)}{\left( Q/1 \right) (\rho_{sd}/10^{-10} \ {\rm kg m}^{-3} )\left( \varv_{pg}(0)/1 \ {\rm ms}^{-1}\right)} 
\end{split}
\label{eq:t_a2}
\eeq
with $Q$ the sticking coefficient: the probability that a dust grain will stick to the particle, $\rho_{sd}$ the spatial density of the dust, $\varv_{pg}(0)$  the speed of the particle relative to the dusty gas at $t=0$ and ${\rho}_p$ is the average mass density of the particle. 

Putting aside the problem that equation~(\ref{eq:Delta_r_th2}) does not explain the non-zero constant $c$ in the observed rim thickness (equation (\ref{eq:Del_r})), we can use the observed result that $\Delta r \approx 0.1a_p(0)$ to show that $t \approx 0.095t_a$. So, all the rimmed particles within, \eg, the CM Murchison carbonaceous chondrite had to accrete onto the Murchison parent body (or be isolated from the dusty gas) at approximately the same time of approximately 360 years after formation - using the scaling parameters in equation (\ref{eq:t_a2}).

\cite{1998Icar..134..180M} noted that this simultaneity requirement is a problem. To this end, they suggested that FGR formation may have exhausted the local dust supply and so we have:

\subsection{Variable spatial mass density of dust}

In this scenario, the macroscopic particles sweep up the dust in the immediate neighbourhood of the particles and thereby limit the growth of the rims. As derived in \cite{1998Icar..134..180M}, the rim thickness for such a scenario is

\beq
\Delta r = a_p(0)\left(\left(\frac{\rho_{sp0}+\rho_{sd0}}{\rho_{sp0}+\rho_{sd0}e^{-t/\tau_a}}\right)^{1/3} - 1\right) \ ,
\label{eq:Dr_gas_depletion}
\eeq
where

\beq
\begin{split}
	\tau_a & = \frac{t_{a}}{3\left(1 + \rho_{sp0}/\rho_{sd0} \right)} \\
	& \approx  630 \ {\rm yr} \
	\frac{({\rho}_p/3 \ {\rm g cm}^{-3}) \left(a_p(0)/1 \ {\rm mm} \right)}{\left( \frac{Q}{1}\right) \left(\frac{\rho_{sdo}}{10^{-10} \ {\rm kg m}^{-3}} \right)\left( \frac{\varv_{pg}(0)}{1 \ {\rm ms}^{-1}}\right)\left(1+\left(\left(\frac{\rho_{sp0}}{\rho_{sd0}}\right)/1\right)\right)} \ ,
\end{split}
\label{eq:tau_a}
\eeq

here $\rho_{sp0}$ and $\rho_{sd0}$ are the initial spatial densities of the macroscopic particles (\eg, chondrules, CAIs, \&c) and dust particles, respectively.

This is an elegant solution, but again, equation~(\ref{eq:Dr_gas_depletion}) does not account for the non-zero constant in the observed rim thickness (equation (\ref{eq:Del_r})). As with  equation~(\ref{eq:Delta_r_th1}), and as \cite{1998Icar..134..180M} acknowledge, this theory still requires additional information regarding the solar proto-planetary disk (SPPD). This theory only works if the dust replacement time at a particular location in the disk is long relative to $\tau_a$ - the "sweep-up" time scale for the macroscopic particles to accrete most of the free dust. So the simultaneity problem is not completely solved as one has to explain why the dust sweep-up timescale should be significantly shorter than the dust replacement timescale.

There has been some work in placing this theory within the context of the SPPD , \eg, \ibid and \cite{2011Icar..211..876C}, but issues such as micro-chondrule formation, low porosity and the fine-grained composition of FGRs remain unexplained by these theories. In the next section, by using the KDA process, we develop a preliminary theory for understanding all these phenomena in the context of a highly kinetic FGR formation process.

\subsection{Kinetic FGR formation}
\label{sec:kineticFGRformation}

In this model, we assume that the initial unrimmed chondrule/CAI or similar macro-sized particle (macro-particle) enters the dusty gas at a relatively high speed. This speed could be of order metres per second to kilometres per second. This situation may arise due to shock waves moving through the protoplanetary disk, so the micron-sized dust is more readily entrained with fast moving gas relative to larger macro-particles (\cite{1998Sci...280...62C,2005ASPC..341..821B}). Alternatively, chondrules may be produced by the collisions between planetsimals and were subsequently launched at high speed into the surrounding dusty gas (\cite{2015Natur.517..339J}). Finally, CAIs may have been formed near the proto-Sun and flung out at high speed only to subsequently re-enter the SPPD at distances significantly further away from the proto-Sun (\cite{2016MNRAS.462.1137L}). This idea may also be true for chondrules or at least reprocessed chondrules (\cite{1996cpd..conf..285L}). Other scenarios may be possible, these are only given as illustrative examples.

\begin{figure}
	\includegraphics[width=\columnwidth]{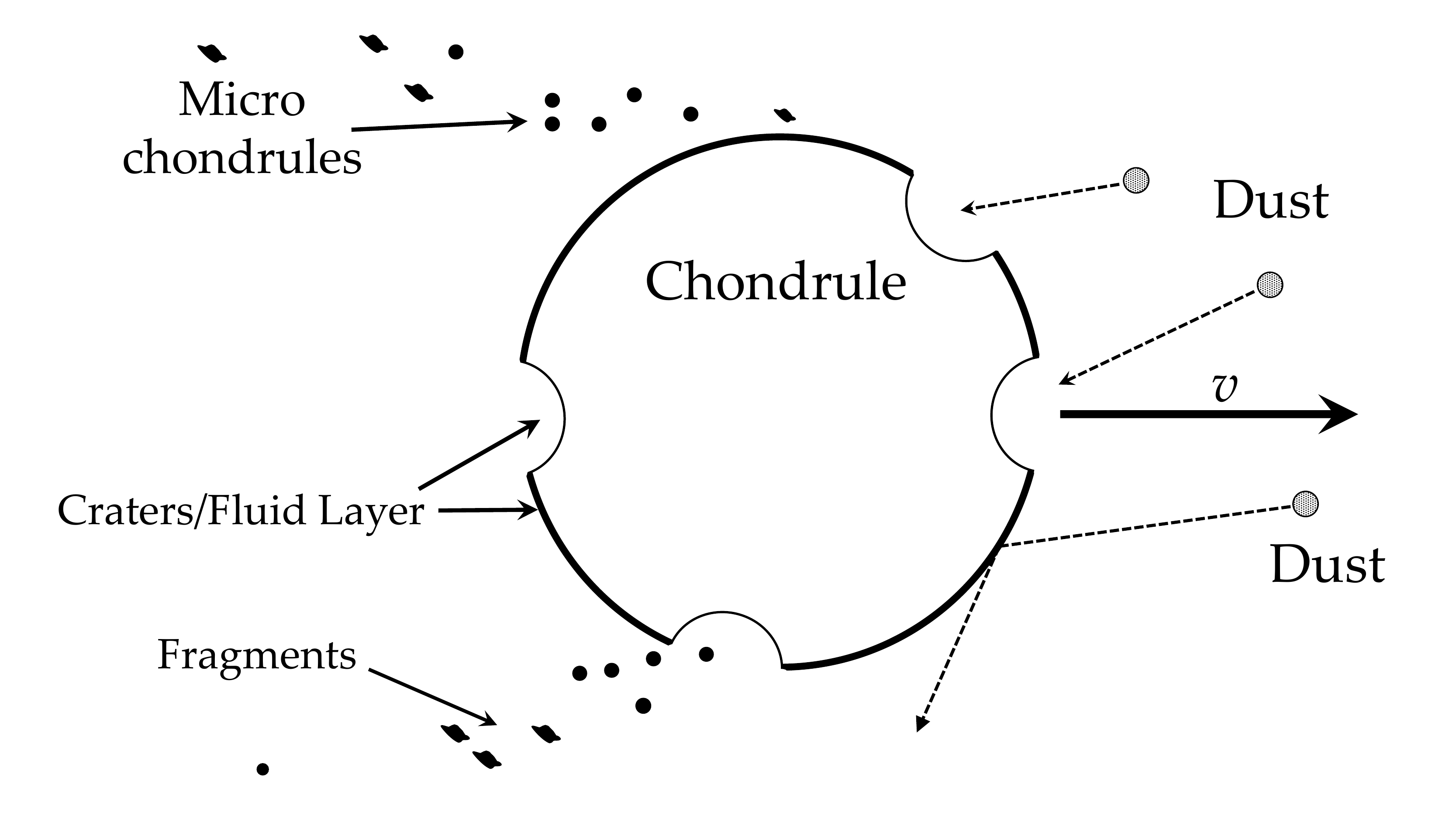}
	\caption{A macro-particle, in this case a chondrule, is travelling through a dusty gas at a speed in excess of the erosion speed. As a consequence, the macro-particle is being eroded by the high speed impact between the macro-particle and the dust. This causes pitting of the particle's surface with the particle producing rock fragments and, if the surface is semi-molten, melt droplets, which we call micro-chondrules.}
	\label{fig:vgve}
\end{figure}

In Figure \ref{fig:vgve}, we show the case where the relative speed between the chondrule/CAI particle is in excess of the erosion speed. In this case, the surface of the particle is pitted  due to the impact of the high speed dust grains with fragments of chondrules being produced by dust erosion. if the surface is semi-molten due to external heating then microchondrules may also be produced by this process. We note that collisions with other chondrules may also be part of the erosive process (\eg, \cite{2019MNRAS.483.4938U}), but we do not consider interchondrule collisions in this paper.

\subsubsection{Initial speed greater than the erosion speed: $\varv_0 \geq \varv_{er}$}

We start with the macro-particle entering the dusty gas with an initial radius of $a_0$ and initial speed of $\varv_0$. We assume that the initial speed is higher than the erosion speed, $\varv_{er}$. The macro-particle will be subject to gas/dust drag and the macro-particle will eventually slow down and reach the erosion speed, $\varv_{er}$. During this time the macro-particle will be subject to partial or complete erosion, where, in this case, we assume it will obtain a non-zero radius of $a_{er}$. Noting the formulae developed in  \ref{sec:KDA}, $a_{er}$ is given by

\beq
    a_{er} \approx a_0
	  -\frac{Q_{er}}{3}a_0 \ln \left( \frac{E+DM_0}{E+DM_{er}}  \right) \ ,
	\label{eq:a_erII}
\eeq
here $Q_{er}$ is the erosion coefficient, $E$ is a numerical factor ($= \frac{8(a+\pi/8)}{3\sqrt{\pi}}$), $D$ is the dust to gas mass ratio, $M_0 = \varv_0/\varv_T$ is the initial Mach number, while $M_{er} = \varv_{er}/\varv_T$ is the erosion Mach number.

\subsubsection{ $\varv_{eq} \leq \varv < \varv_{er}$}

Drag forces continue to slow the macro-particle so it enters a regime where the collisions with dust still erode the particle, but some of the colliding dust starts to stick to the macro-particle. As a consequence, a layer of accreted dust, collisionally produced microchondrules and macro-particle fragments builds up on the macro-particle's surface. We set the thickness of this layer to the average size of a compressed dust particle and we denote the thickness of this layer by $\delta$. We presume that the macro-particle is much larger than the dust particles so $\delta$ is somewhat independent of the macro-particle size. This interface layer of collisionally compressed dust between the chondrule and the subsequent FGR will be dependent on the dust particle's initial size, speed of impact, density and composition.

\begin{figure}
	\includegraphics[width=\columnwidth]{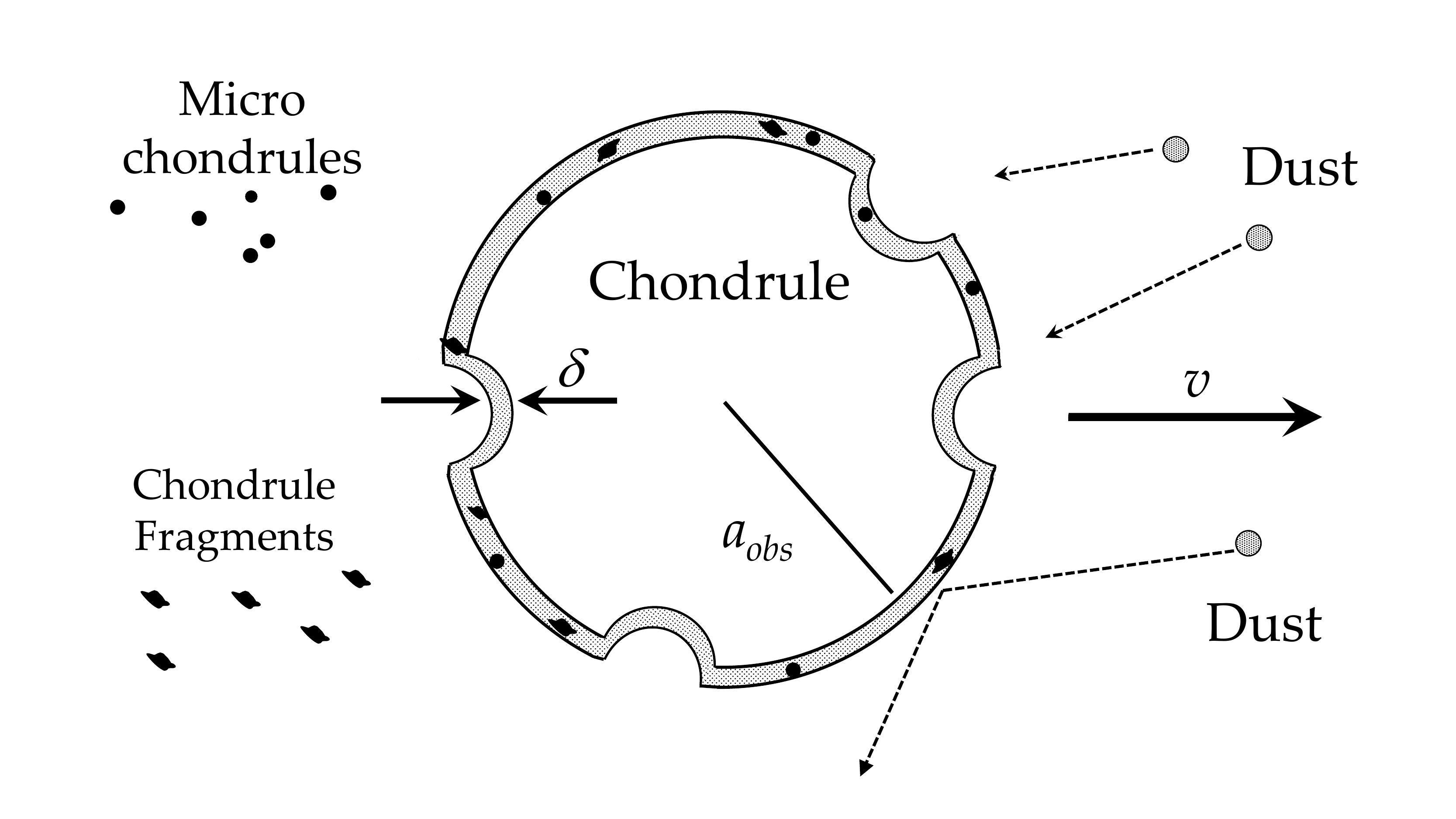}
	\caption{In this case, the chondrule is travelling with a speed relative to the dusty gas in the range $\varv_{eq} \leq \varv < \varv_{er}$. In this speed range, the chondrule is still suffering erosion, but dust is also kinetically accreting onto the surface of the chondrule and within the hollows of the collisionally produced craters. As such, microchondrules and chondrule fragments are captured in an accreted rim of dust of a set thickness $\delta$, where $\delta$ is of order the average size of a compressed dust grain. The $\delta$ depth is semi-independent of the chondrule size, but is dependent on the dust collisonal speed, dust composition and the size of the dust particles}
	\label{fig:delta}
\end{figure}

The macro-particle is no longer subject to erosion once the speed of the macro-particle reaches the speed $\varv_{eq}$. As such, it reaches a final or observed internal radius, $a_{obs}$ given by

\beq
a_{obs} \approx a_{er}
-\frac{Q_{er}}{3}a_{er} \ln \left( \frac{E+DM_{er}}{E+DM_{eq}}  \right) \ ,
\label{eq:a_er}
\eeq
while the thickness of the rim is simply

\beq
\Delta r = \delta \ .
\eeq
So, the final radius of the macro-particle is

\beq
a_p(t_{eq}) = a_{eq} = a_{obs} + \delta \ .
\label{eq:a_p(t_eq)}
\eeq
This situation is schematically depicted in Figure \ref{fig:delta}.

It should be noted that if the initial speed of the macro-particle relative to the dusty gas is initially less than $\varv_{eq}$ then one will not observe significant erosion or micrchondrule/chondrule-fragment formation. However, the $\delta$ factor will still be present if the initial speed was greater than $\varv_{cr}$ as $\delta$ represents the thickness of the first initial layer of dust that adhered to the chondrule or other macro-particle.

\subsubsection{ $\varv_{cr} \leq \varv < \varv_{eq}$}

\begin{figure}
	\includegraphics[width=\columnwidth]{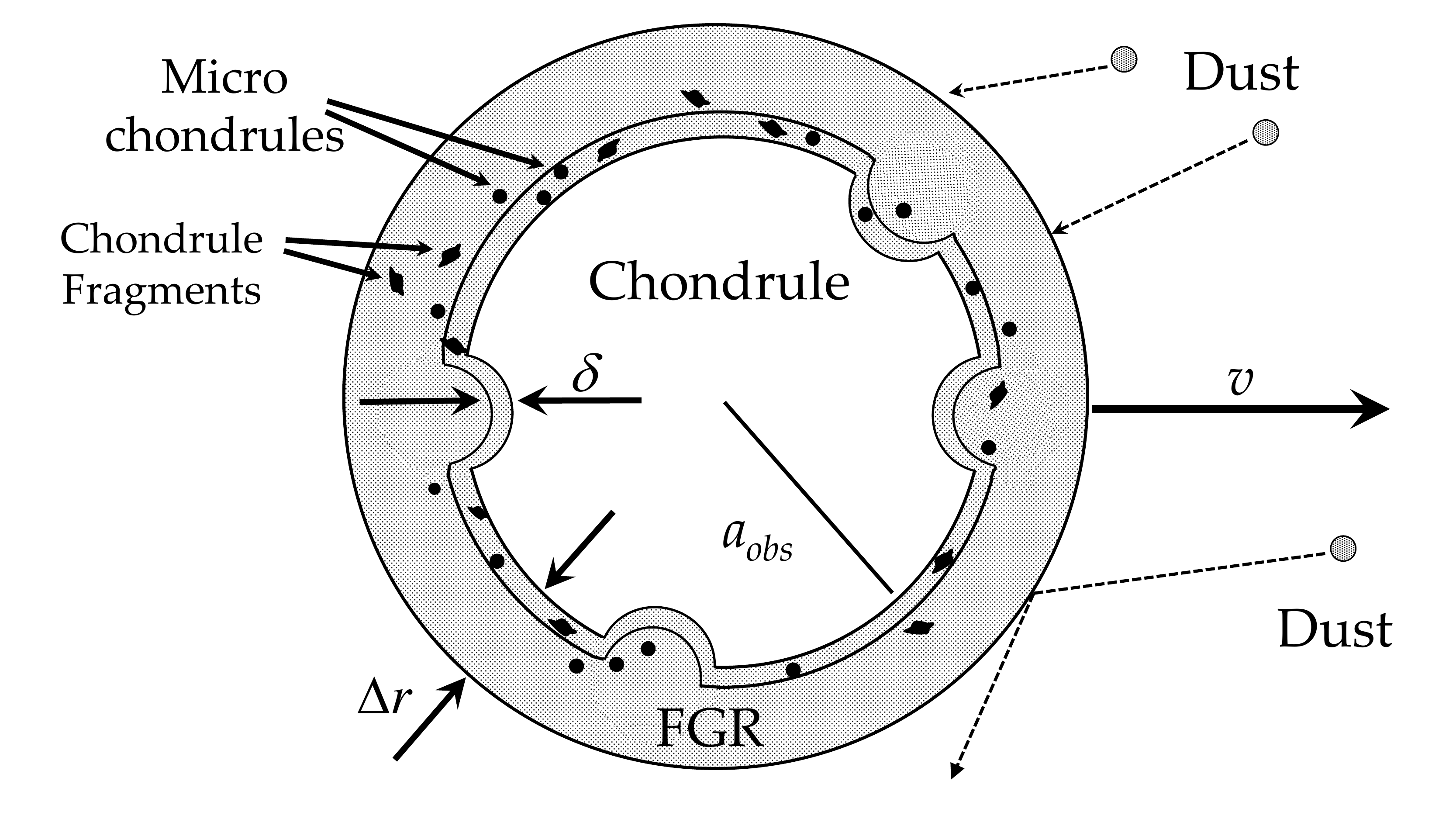}
	\caption{The chondrule is travelling with a speed in the range $\varv_{cr} \leq \varv < \varv_{eq}$. In this speed range, the chondrule is no longer suffering erosion and dust is kinetically accreted onto the surface of the chondrule to form the fine grained rim (FGR). The initial layer of thickness $\delta$ is shown for schematic purposes only and would probably not show up as a separate layer.}
	\label{fig:vltveq}
\end{figure}

The macro-particle has now reached a speed range where the FGR builds up in size and there is no erosion. Once the macro-particle slows to a speed below the critical speed, $\varv_{cr}$ then dust adhesion stops and the macro-particle reaches its final and maximum radius (Figure \ref{fig:vltveq}), $a_f$:

\beq
\begin{split}
a_f & \approx a_{eq} + \frac{Q_{ad}}{3}a_{eq} \ln \left( \frac{E+DM_{eq}}{E+DM_{cr}}  \right) \\
& = a_{obs} + \delta + \frac{Q_{ad}}{3}( a_{obs} + \delta) \ln \left( \frac{E+DM_{eq}}{E+DM_{cr}}  \right)\ ,
\label{eq:a_f}
\end{split}
\eeq
with $Q_{ad}$ the adhesion or sticking coefficient and $M_{cr} = \varv_{cr}/\varv_T$ the Mach number for the critical speed.

The observed rim thickness, $\Delta r_{obs}$ is

\beq
\begin{split}
\Delta r_{obs} & = a_f - a_{obs} \\
&\approx \frac{Q_{ad}}{3} a_{obs}  \ln \left( \frac{E+DM_{eq}}{E+DM_{cr}} \right) \\
& \ \ \ \ \ \ \ \ \ \ \ \ \
+ \delta\left(1+\frac{Q_{ad}}{3}  \ln \left( \frac{E+DM_{eq}}{E+DM_{cr}} \right)\right) \ .
\label{eq:Deltar_obs}
\end{split}
\eeq
Hence the observed rim thickness equation for the Murchison chondrite (equation \ref{eq:Del_r}) is:

\beq
\Delta r_{obs} = K a_{obs} + c \ \approx 0.1 a_{obs} + 26\mu m \ ,
\label{eq:Deltar_obsII}
\eeq
with

\beq
K = \frac{Q_{ad}}{3} \ln \left( \frac{E+DM_{eq}}{E+DM_{cr}} \right)
\label{eq:K_obs}
\eeq
and

\beq
c = \delta \left(1+\frac{Q_{ad}}{3}  \ln \left( \frac{E+DM_{eq}}{E+DM_{cr}} \right)\right) = \delta ( 1 + K)
\label{eq:c_obs}
\eeq

\subsection{Observations and Constraints}

From equation (\ref{eq:Deltar_obsII}), we have for the Murchison chondrite that

\beq
K \approx 0.1
\eeq
and

\beq
c = 26 \ \mu m.
\eeq
This combined with equation (\ref{eq:c_obs}) implies

\beq
\delta = \frac{c}{1+K} \approx 24 \mu m \ .
\eeq
If we can experimentally determine the values for $Q_{ad}$, $\varv_{eq}$ and $\varv_{cr}$ then, from equation (\ref{eq:K_obs}), we should be able to determine a range of values for the dust to gas mass ratio, $D$, via the equation:

\beq
D = \frac{E\left( \exp\left(\frac{3K}{Q_{ad}} \right) - 1\right)}{M_{eq}- M_{cr}\exp\left(\frac{3K}{Q_{ad}} \right)} \ .
\eeq
The requirement that $D > 0$ implies that

\beq
M_{cr} < M_{eq}\exp\left(-\frac{3K}{Q_{ad}}\right) \ .
\eeq
Hence, if we were to use the observed $K \approx 0.1$ and note that current experiments (which may not be appropriate for FGR materials) indicate that $Q_{ad} \approx 0.1$ then

\beq
M_{cr} < M_{eq}\ e^{-3} \approx 0.05M_{eq} \ .
\eeq
It follows that if $\varv_{eq} \approx 300$ ms$^{-1}$ then the critical dust collision speeds for the onset of dust adhesion may be smaller than 15 ms$^{-1}$. As noted previously, \cite{2008ARA&A..46...21B} report a critical speed for adhesion at a speed of 13 ms$^{-1}$ for their "C3" experiments.

	\section{Discussion}

How micron-sized dust in the early Solar System agglomerated into planets is still an open question (\cite{2017MNRAS.467.1984G}). Fine Grained Rims (FGRs) provide one example of how dust aggregation may have occurred. The popular model for the formation of FGRs in the early Solar System is the aggregation of dust grains onto the chondrule/CAI's surface via slow speed collisions of less than a metre per second. The resulting "fractal" or fluffy rim was then compacted due to compressive forces after the chondrule/CAI + rim were incorporated into a planetesimal. This general idea of low speed dust aggregation within the Solar Protoplanetary Disk (SPPD) is physically valid, but the phase space of dust collisional behavior is large and there are other experimentally proven pathways for dust aggregation. 

We argue that the slow-speed FGR formation model does not explain the highly compacted/low porosity structure of FGRs, the fine-grained nature of FGRs, the energetic formation of micro-chondrules and chondrule fragments embedded in FGRs, the eroded/cratered surface of chondrules underneath a FGR, the linear relation between rim thickness and the radius of the chondrule and the non-zero constant in this linear relation, where the latter property shows that FGR formation had an effect on the chondrule surface that was independent of the chondrule's size. Indeed, the non-zero constant may represent the diameter of the compacted dust grains that formed the initial layer of the FGR.

Instead, we suggest that all these properties, and more, can be understood in the context of FGR formation being due to the relatively high speed collisions of dust with CAIs and chondrules, where the micron-sized dust was fragmented upon collision and the subsequent fragments underwent vacuum bonding to produce the rim. A similar process, called Aerosol Deposition, is used in the ceramics and semi-conductor industries. We denote this process, as applied to Solar System, as Kinetic Dust Aggregation (KDA).

As discussed in \cref{sec:kineticFGRformation}, Chondrules and CAIs may have been produced or transported via energetic processes, \eg, planetesimal collisions, shock waves and/or jet flow ejection/processing from the inner to outer regions of the SPPD.  In all these cases, the macro-particles entered the dusty gas of the SPPD at high speed. These high speed macro-particles were subject to gas and dust drag and were brought to rest relative to the dusty gas. If the initial speeds were high enough then the chondrule/CAI were subject to erosion and heating due to gas drag and the "sand-blasting" effect of high speed dust collisions. Chondrule fragments were a necessary byproduct of such erosion, which may have also produced micro-chondrules - if there were some external heat source that remelted the chondrule surface. Subsequent gas drag and slower speeds would have allowed the formation of the FGR via the KDA process. Finally, the rimmed chondrule/CAI speed dropped below the critical speed such that the KDA process stopped and the final size of the rimmed chondrule/CAI was obtained.

This energetic rim formation process can provide a basis for understanding the compact and fine-grained nature of FGRs, the eroded chondrule surfaces found under FGRs plus the production and incorporation of chondrule fragments and micro-chondrules within FGRs. The linear relationship between FGR thickness and the size of the enclosed chondrules can be intuitively understood by noting that a chondrule moving at high speed relative to a dusty gas will come to rest when the chondrule has encountered a dust and gas mass approximately equivalent to the mass of the chondrule. 

A similar result should apply to FGRs around CAIs, where we would expect to see some proportionality between the thickness of FGR rims and the size of CAIs. Although the irregular shapes of many CAIs may make this measurement  difficult to accomplish. 

The KDA model also suggests that if a macroparticle were subject to more than one, sufficiently high speed traverse through a dusty gas then we might expect multiple fine grained rims to form on the macroparticle.  \citet{1992GeCoA..56.2873M} found that most of the rims in the Kivesvaara meteorite were multi-layered. Although, such multi-layered rims might have also formed from just one high speed event, where the macro-particle traversed regions of different types of dust. 

There is evidence that a significant proportion of chondrules were reheated and recycled (\cite{2017SciA....3E0407B}). So it might be the case that multiple FGRs on a chondrule were formed, but erosion might have removed the previous FGR or reheating caused the prior FGR to melt and produce a multilayered chondrule. \citet{2013M&PS...48..445R} discusses a multi-layered chondrule that appears to have four separate layers, which may be an example of this FGR reprocessing. 

Multi-layered FGRs may also be an experimental test for the KDA model. For example, proper analytic, computational and experimental modelling may show that the KDA process will produce finer grained rims in the interior of the rim near the surface of the macro-particle with the grain size increasing as one approaches the outer surface of the rim. This may arise due to the higher speed of the colliding dust grains at the start of the rim formation process relative to the end of the process where the macro-particle has slowed down and dust collision speeds are much lower. Similarly, with chemical reactions, we might expect higher temperature reactions at the most interior regions of the rim relative to the surface of the rim. This, of course, would also depend on the composition and size of the impacting dust particles. 

The KDA process may provide a tentative, indirect explanation for why some chondrules in some chondrites don't have FGRs. The first, obvious answer is that the chondrules may have traversed a gas with very low dust content. If there is little or no dust relative to the number of chondrules then one cannot have FGR formation. Putting aside this trivial answer, another answer arises due to the critical speed for the onset of dust adhesion in the KDA process. If the chondrule's initial speed was lower than the critical speed for the adhesion of dust onto the chondrule/macro-particle's surface then no FGR formation would have occurred. According to the traditional model of FGR formation, slow relative speed of the macro-particle to the dusty gas is the requirement for FGR formation. The KDA process is the opposite. One requires a relatively high initial speed under vacuum conditions to allow Kinetic Dust Aggregation to occur. This leads to the final potential constraint for FGR formation in that one also requires vacuum conditions. If some chondrules formed in a relatively high pressure environment then the dust may not have adhered, even if the critical collision speeds had been obtained 

	\section{Conclusions}

In this study, we introduced the Kinetic Dust Aggregation (KDA) model for fine grained rim formation around chondrules and CAIs. KDA is based on the observation that coatings which are qualitatively similar to fine grained rims can be produced when micron-sized particles impact a substrate at speeds of order 100 ms$^{-1}$ under vacuum conditions. The coatings are formed due to the fragmentation of the micron-sized particles, which consumes the impact kinetic energy and allows the sub-micron sized fragments to bond to each other and the substrate due to the subsequent increase in contact surface area between the resulting fragments.

This vacuum bonding can only arise within a range of dust impact speeds between a critical onset speed for adhesion and a speed which produces the onset of full erosion. For this process to produce fine grained rims, we require that chondrules and CAIs enter the dusty gas of the Solar Protoplanetary Disk (SPPD) at relatively high speeds. This high initial speed may, for example, be due to shock waves in the SPPD (\cite{2005ASPC..341..821B}), chondrule formation due to planetesimal collisions (\cite{2015Natur.517..339J}) and/or high speed transport from the inner to outer regions of the SPPD due to outflow ejection processes in the very inner disk regions adjacent to the protoSun (\cite{1996cpd..conf..285L},\cite{2016MNRAS.462.1137L}).

At high enough initial speeds, the chondrules/CAIs are eroded by dust collisions. This erosion process will produce chondrule fragments and if the surface of the chondrule is molten may also induce micro-chondrule formation. As the chondrules/CAIs slow then the colliding dust grains  produce the FGRs until the chondrules and CAIs reach low enough speeds and the KDA process turns off. This is a completely different model from the standard slow speed collision model for dust aggregation in protoplanetary disks, where it is assumed that most dust aggregation occurs at collision speeds of less than approximately a metre per second. 

The KDA process naturally explains the observed low porosity and fine-grained nature of FGRs, it provides a basis for understanding the eroded surfaces of chondrules, the presence of chondrule fragments and micro-chondrules in FGRs and naturally produces the liner relationship between FGR thickness and the radius of the inner object, where the non zero constant in the linear relation  may be due to the average size of the compressed dust grains that produced the initial layer between the FGR and the macro-particle.

Fine grained rim formation is one of the few examples of dust aggregation that has survived in the meteoritic record to the present day. The energetic nature of fine grained rim formation is illustrative of how high speed collision of dust particles can produce stable, durable aggregates. As such, Kinetic Dust Aggregation may be a dust aggregation process that is applicable to not only the early Solar System, but also protostellar disks in general.

	\section*{Acknowledgements}
	
	The author acknowledges the support from the Swinburne University of Technology and Professor Sarah Maddison. The author wishes to thank and acknowledge, with gratitude and appreciation, the two anonymous referees and Dr Pia Friend (Universit\"at zu K\"oln) whose constructive suggestions and comments definitely improved the paper.
	
	
	\newpage
	
\section*{References}

\bibliography{paper_finerims}

	
	
	\appendix
	
    \newpage
	
	\section{Rim Formation with Constant Spatial Mass Density of Dust}
	\label{sec:const_density}
	
	In this derivation of accretion rim thickness, we assume that a macroscopic particle moving through a dusty gas, where the spatial mass density of the dust is kept constant. To derive the rim thickness of the accreted dust, we start with the equation for the rate of growth in size for such a macroscopic particle:

\beq
\dl{a_p}{t} \approx Q\frac{\rho_{sd}\varv_{pg}}{4 {\rho}_p} \ ,
\label{eq:accretion}
\eeq

Here $a_p$ is the radius of the macroscopic particle, $t$ is time, $Q$ is the sticking coefficient: the probability that a dust grain will stick to the particle, $\rho_{sd}$ is the spatial density of the dust, $|\varv_{pg}|$ is the speed of the particle relative to the dusty gas and ${\rho}_p$ is the average mass density of the particle. This is an approximate equation, because the derivation implicitly assumes that ${\rho}_p$ is approximately constant with time.
We note that

\beq
\rho_{sd} = \overline{m}_d n_d \ ,
\label{eq:rho_sd}
\eeq
with $\overline{m}_d$ is the average mass of a dust grain and $n_d$ is the number density of dust grains. While

\beq
\varv_{pg} = |\vec{\varv}_p - \vec{\varv}_g| \ ,
\label{eq:varv_pg}
\eeq	
with $\vec{\varv}_p$ the average velocity of the particles and $\vec{\varv}_g$ the average velocity of the gas.

\cite{1998Icar..134..180M} consider the case where the dust is well coupled to the gas (\ie, the drag time-scale of the dust to the gas, $\tau_{dg}$, is small relative to the orbital period) and so

\beq
\varv_{pg} \approx \frac{\tau_{dg}}{\rho_g} \left|\pl{p_g}{r}\right| \ ,
\label{eq:varv_pgII}
\eeq
where $\rho_g$ is the mass density of the gas, $p_g$ is the gas pressure and $r$ is the cylindrical radial distance from the centre of the Sun.

In the Epstein drag regime (where $\varv_{pg} \ll C_s$ with $C_s$ the sound speed and $a_p \ll \lambda_g$, $\lambda_g$ the mean free path of the gas) the drag time-scale has the form

\beq
\tau_{dg} = \frac{\sqrt{\pi}}{2(1+\pi/8)}\frac{a_p \rho_p}{\varv_T \rho_g} \ ,
\label{eq:tau_dg}
\eeq
the unusual coefficient in the above equation arises from the Epstein drag coefficient equation derived in the Appendix of \cite{2000Icar..143..106L}, while $\varv_T$ is the thermal gas speed:

\beq
\varv_T \approx \sqrt{\frac{2k_BT_g}{\overline{m}_g}} \ ,
\eeq
 $k_B$ is the Boltzmann constant, $T_g$ is the temperature of the gas and $\overline{m}_g$ is the mean molecular mass of the gas.

Combining equations (\ref{eq:varv_pgII}) and (\ref{eq:tau_dg}) gives

\beq
\varv_{pg}(t) = \frac{a_p(t)}{a_p(0)}\varv_{pg}(0) \ ,
\eeq
which allows us to rewrite equation (\ref{eq:accretion}) as

\beq
\dl{a_p}{t} \approx \frac{a_p}{t_a} \ .
\eeq
So the rim thickness, $\Delta r$, is

\beq
\Delta r = a_p(0)(e^{t/t_a} - 1) \ .
\label{eq:Delta_r_th1}
\eeq
The accretion time-scale, $t_a$, has the form

\beq
\begin{split}
t_a & = \frac{4{\rho}_pa_p(0)}{Q\rho_{sd} \varv_{pg}(0)} \\
& \approx 3,800 \ {\rm yr} \
\frac{({\rho}_p/3 \ {\rm g cm}^{-3}) \left(a_p(0)/1 \ {\rm mm} \right)}{\left( Q/1 \right) (\rho_{sd}/10^{-10} \ {\rm kg m}^{-3} )\left( \varv_{pg}(0)/1 \ {\rm ms}^{-1}\right)}
\end{split}
\label{eq:t_a}
\eeq

	\newpage
	\section{Kinetic Dust Accretion Equation}
	\label{sec:KDA}

	Consider the drag equation for a macroscopic particle (\eg, a chondrule, CAI, \&c) moving through a dusty gas:

	\beq
	\dl{\varv_{pg}}{t} = -\frac{3}{8 a_p \rho_p} \left(\rho_g C_{Dg} + \rho_{sd} C_{Dd}  \right)\varv_{pg}^2 \ ,
	\label{eq:drag}
	\eeq
	where $C_{Dg}$ is the gas drag coefficient and $C_{Dd}$ is the dust drag coefficient. The other variables in this equation have been defined in Section \ref{sec:FGR_theory}. The drag coefficients can be determined from \cite{2000Icar..143..106L}:

	\beq
	C_{Dd} = 2
	\label{eq:C_Dd}
	\eeq
    and for the Epstein Drag regime
  
    \beq
    C_{Dg} \approx \frac{16(1+\pi/8)\varv_T}{3\sqrt{\pi}\varv_{pg}} \ .
    \label{eq:C_Dg}
    \eeq
    Combining the above equations gives
  
    \beq
    \begin{split}
    \dl{\varv_{pg}}{t} &= -\frac{2\rho_g\left(1 + \pi/8\right) \varv_T }{a_p \rho_p \sqrt{\pi}}\varv_{pg}  - \frac{3 \rho_{sd}}{4 a_p \rho_p} \varv_{pg}^2 \\
    &= -A\varv_{pg}-B\varv_{pg}^2 \,
    \label{eq:drag2}
    \end{split}
    \eeq
    which has the solution
 
    \beq
    \varv_{pg}(t) = \frac{A\varv_0}{A e^{At}-B\varv_0(1 - e^{At})} \ ,
    \label{eq:v_analytic}
    \eeq
    with $\varv_0 = \varv_{pg}(0)$,
  
    \beq
    A = \frac{2\rho_g\left(1 + \pi/8\right) \varv_T }{a_p(0) \rho_p \sqrt{\pi}} \ ,
    \eeq
    and
   
    \beq
    B = \frac{3 \rho_{sd}}{4 a_p(0) \rho_p} \ .
    \eeq

    Here $A$ and $B$ are constants since we have made the approximation of setting the value of the particle radius, $a_p$, to the initial particle radius. Using equations (\ref{eq:accretion}) and (\ref{eq:v_analytic}), we can write the accretion equation as:
    
    \beq
    \dl{a_p}{t} \approx Q\frac{\rho_{sd}}{4 {\rho}_p}\left(\frac{A\varv_0}{A e^{At}-B\varv_0(1 - e^{At})} \right) \ .
    \label{eq:dapdt}
    \eeq
    Setting $t_i$ as the initial time and $t_f$ the final time,
    we can integrate equation (\ref{eq:dapdt}) and obtain
    \beq
     \Delta r = \int_{t_i}^{t_f}\dl{a_p}{t}dt = a_p(t_f)-a_p(t_i) \equiv a_f - a_i 
    \eeq
   and
   \beq
  \int_{t_i}^{t_f}\left(\frac{A\varv_0}{A e^{At}-B\varv_0(1 - e^{At})} \right)dt =\frac{1}{B}\int_{u_i}^{u_f}\frac{du}{u} = \frac{1}{B}\ln\left(\frac{u_f}{u_i}\right),
   \eeq
   where
   \beq
   u= A - B\varv_0(e^{-At}-1); \ u_f=u(t_f) \ \& \ u_i=u(t_i).
   \eeq
   So, integrating equation (\ref{eq:dapdt}) gives
    \beq
    \Delta r =  \frac{Qa_i}{3}\ln \left( \frac{A - B\varv_0(e^{-At_f}-1)}{A - B\varv_0(e^{-At_i}-1)}\right) 
    \ ,
    \label{eq:Delr_analytic1}
    \eeq
    Using equation (\ref{eq:v_analytic}), we can write
    \beq
    A - B\varv_0(e^{-At}-1)=\frac{A\varv_0e^{-At}}{\varv_{pg}},
     \label{eq:v_analyticII}
    \eeq
    Combining equations (\ref{eq:Delr_analytic1}) and (\ref{eq:v_analyticII}) gives
    \beq
       \Delta r = \frac{Qa_i}{3}\ln \left( \frac{\varv_i}{\varv_f} e^{-A(t_f-t_i)} \right)
    \eeq
    $a_i = a_p(0)$ is the initial radius of the macroscopic particle, $a_f$ the final radius, $\varv_i = \varv_0$ with $\varv_f = \varv(t_f)$.
	
	Using equation (\ref{eq:v_analytic}), we can solve for $t$:
	
	\beq
	t-t_i = \frac{1}{A} \ln \left(\frac{\varv_i (A + B \varv)}{\varv (A + B \varv_i)}\right) \ .
	\eeq
	So

	\beq
	t_f - t_i = \frac{1}{A} \ln \left(\frac{\varv_i (A + B \varv_f)}{\varv_f (A + B \varv_i)}\right) \ ,
	 \label{eq:timediff}
	\eeq
	which implies that

	\beq
	e^{-A(t_f-t_i)} = \frac{\varv_f (A + B \varv_i)}{\varv_i (A + B \varv_f)}
	\label{eq:timediff2}
	\eeq
	and equation (\ref{eq:Delr_analytic1}) becomes
	
	\beq
	\Delta r = \frac{Qa_i}{3}\ln \left( \frac{A + B \varv_i}{A + B \varv_f}  \right)
	=  \frac{Qa_i}{3}\ln \left( \frac{E+DM_i}{E+DM_f}  \right) \
	\label{eq:Delr_analytic2}
	\eeq
	or
	
	\beq
	a_f \approx a_i + \frac{Qa_i}{3}\ln \left( \frac{E+DM_i}{E+DM_f}  \right) \
	\label{eq:Delr_analytic3}
    \eeq
	with
	\beq
	E = \frac{8(1+\pi/8)}{3\sqrt{\pi}} \approx 2.0953 \ ,
	\eeq
	\beq
	D = \frac{\rho_{sd}}{\rho_g}
	\eeq
	$D$ being the dust to gas ratio,

	\beq
	M_i = \varv_i/\varv_T
	\eeq
	and

	\beq
	M_f = \varv_f/\varv_T  \ ,
	\eeq
	$M$ being the Mach number.
	
	As an example for a rim formation timescale, we can set $\rho_g \approx 10^{-10}$ kgm$^{-3}$, $\varv_T \approx 1$ kms$^{-1}$, $a_p(0) \approx 10^{-3} $m, and $\rho_p \approx 3000$ kgm$^{-3}$. These values imply that
	\beq
	\frac{1}{A} = \frac{a_p(0) \rho_p \sqrt{\pi}}{2\rho_g\left(1 + \pi/8\right) \varv_T}
    \approx 0.6 \ \frac{\left(\frac{a_p(0)}{1 {\ \rm mm}}\right) \left(\frac{\rho_p}{3 {\ \rm g cm}^{-3}}\right)}{\left(\frac{\rho_g}{10^{-10} \ {\rm kgm}^{-3}}\right)\left(\frac{\varv_T }{1 {\rm \ kms}^{-1}} \right)} \ {\rm yrs} \ .
	\eeq
	To calculate the $B$ factor, we note that the dust to gas mass ratio in the interstellar medium is very approximately 1\% (\citet{1978ppim.book.....S}), so we set $\rho_{sd} \approx  0.01\rho_g$ as a first approximation and we obtain
	\beq
	B = 2.5\times10^{-13} \frac{\left(\frac{\rho_{sd}}{10^{-12} \ {\rm kgm}^{-3}}\right)}{\left(\frac{a_p(0)}{1 {\ \rm mm}}\right)\left(\frac{\rho_p}{3 {\ \rm g cm}^{-3}}\right)} \ {\rm m}^{-1} \ .
    \eeq
    Substituting these values with $\varv_i \approx $ 100 ms$^{-1}$ and $\varv_f \approx$ 10 ms$^{-1}$ into equation (\ref{eq:timediff}) implies that
    \beq
    \Delta t = t_f - t_i \approx 1.4 \ {\rm yrs} \ .
    \eeq
    So with these given representative values, the timescale for KDA rim formation is of order years.

	\label{lastpage}
\end{document}